# Design, implementation and application of the Marchenko plane-wave algorithm

*Jan Thorbecke,* *Mohammed Almobarak,† Johno van IJsseldijk,‡
Joeri Brackenhoff,§ Giovanni Meles¶and Kees Wapenaar ‖*

**Abstract**

The Marchenko algorithm can suppress the disturbing effects of internal multiples that are present in seismic reflection data. To achieve this, a set of coupled equations with four unknowns is solved. These coupled equations are separated into a set of two equations with two unknowns using a time window. The two unknown focusing functions can be resolved by an iterative or direct method. These focusing functions, when applied to reflection data, create virtual point-sources inside the medium. Combining individual virtual point-sources into a plane-wave leads to an efficient computation of images without internal multiples. In this study the internal multiples are eliminated in a redatuming step which is part of the imaging algorithm. To use the Marchenko algorithm with plane-wave focusing functions, the time window that separates the unknowns must be adapted. The design of the plane-wave Marchenko algorithm is explained and illustrated with numerically modeled and measured reflection data.

## Introduction

Seismic imaging is a technique to image geological structures in the subsurface of the earth from reflected wavefields measured at the surface of the earth. The measured wavefields usually originate from human-activated and controlled sources such as air-guns or vibrating plates. In passive seismic methods the source of the wavefield can originate from earthquakes, ocean waves, or uncoordinated human activities as traffic. The primary reflection of a geological structure, large and strong enough to be detected by a propagating wavefield, is of main interest and is used to compute an image of the subsurface. At each geological structure, wavefields are partly reflected upward and partly transmitted further downward. Between two strong reflecting structures, the wavefield can bounce up and down multiple times and generate so called internal multiples. These multiple reflections are also measured by geophones at the surface and difficult to distinguish from primary reflections. In the imaging step the reflections are migrated from time to depth and construct an image of the subsurface. If multiple reflections are not recognized as such, they will get imaged being primary reflections at wrong depths. These falsely imaged multiples distort the actual image of the subsurface; the distorted image contains much more (ghost) structures that are positioned along with the primary reflections. Therefore, in seismic imaging, it is important to recognize these multiple reflections, and if possible, remove them from the computed image. This removal can be performed at different stages of the processing scheme to construct an image of the subsurface. The internal multiples can be directly removed from the measured reflection data, in the redatuming step or after the imaging step. For removal after the imaging step, a computed prediction of imaged multiples is subtracted from the image to obtain an image without multiples. In this paper, we discuss a method for removing internal multiples during the redatuming step.

Besides internal multiples that are reflected between boundaries within the subsurface, there are also free-surface-related multiples. These multiples are generated by reflections from upcoming waves that bounce back into the subsurface by the surface of the earth. Free surface multiples are not considered in this paper. They are assumed to be removed prior to the removal of the internal multiples.

The Marchenko algorithm can eliminate internal multiples from seismic reflection data (Slob et al. (2014); Behura et al. (2014)). In this algorithm the up- and down-going focusing functions, with a focal-point in the subsurface, are key to the method. The goal of the Marchenko method is to retrieve these up- and down-going parts of the focusing functions from the reflection data by solving a coupled set of the so-called Marchenko equations. This set of equations can be solved by iterative methods (Wapenaar et al. (2014b); Thorbecke et al. (2017)), or a direct method (van der Neut et al. (2015a); Ravasi (2017)). The Marchenko method has found many different applications, ranging from redatuming for time-lapse monitoring (van IJsseldijk et al. (2022)), adaptive subtraction of Marchenko estimated multiples (Staring et al. (2018)), homogeneous Green's function retrieval (Brackenhoff et al. (2019)), and direct multiple elimination on reflection data (Zhang and Slob (2020)). In this paper, a particularly efficient application of the Marchenko method for imaging by plane-waves is highlighted, and the implementation details of this method are discussed in more detail.

Meles et al. (2018) show that besides focal-points, focal-planes can also be solved by the Marchenko equations. Meles et al. (2020) build on the Marchenko Multiple Elimination (MME) method proposed by Zhang and Slob (2019) and introduce the plane-wave MME method. The major advantage of the plane-wave-based Marchenko method is that with a minimal effort of one Marchenko run (for a single plane-wave), for each depth level (or time instant for MME), a multiple free image can be built up. Especially in 3D applications, the plane-wave Marchenko method is computationally efficient for building an internal multiple-free image (Brackenhoff et al. (2022)). A single plane-wave can be sufficient to build an accurate image if the subsurface interfaces are near-flat. Multiple plane-waves, with different illumination angles, are needed for subsurface interfaces with varying dips, or geologically complex structures, to illuminate the subsurface properly (Almobarak (2021)). The Marchenko algorithm for the focal-plane method is similar to the focal-point algorithm. The initial point-focusing function is in the plane-wave algorithm replaced by a time-reversed direct plane-wave response. The main difference lies in the choice of the time windows to separate the Green's functions from the focusing functions. The minimum conditions to hold for a plane wave are the same as for a point source; a separation in time can be made between the direct and later arrivals. In this paper, we discuss in detail how these time windows are adapted for the Marchenko plane-wave method, discuss the implementation aspects, and illustrate the method with applications on numerically modeled and field data.

The software accompanied by this paper contains scripts and source code to reproduce all the numerical examples presented in this paper. The code can also be found in its GitHub repository (Thorbecke et al. (2017); Thorbecke and Brackenhoff (2023)), where the most recent updated version and the latest developments are available. To reproduce the figures and perform a few pre- and post-processing steps, Seismic Unix (Stockwell and Cohen (2016)) is required.

## Theory

The Marchenko method is introduced by two coupled equations that contain four unknown fields (up- and downgoing focusing functions and Green's functions) we would like to retrieve. These fields enable us to suppress internal multiples by using the up- and downgoing focusing functions to redatum seismic reflection data from the acquisition surface $\mathbb{S}_0$ to the focal level(s) in the subsurface. In this redatuming step, the internal multiples of the overburden are suppressed. The seismic reflection data is recorded at acquisition boundary $\mathbb{S}_0$. The reflection response $R(\mathbf{x}_R, \mathbf{x}_S, t)$, a scaled version of the Green's function without the direct arrival (Wapenaar et al. (2012)), is measured with sources and receivers positioned at $\mathbf{x}_S$ and $\mathbf{x}_R$ on this boundary. The recording time is denoted by $t$. This reflection response does not contain free-surface related multiple reflections neither a source wavelet. Hence, pre-processing steps are required to remove the free-surface multiples and the direct arrival, and to deconvolve the wavelet from the measured reflection data.

The up- and downgoing parts of the focusing functions $f_1^-$ and $f_1^+$ are used to define a relation between the decomposed Green's functions $G^{-,+}$ and $G^{-,-}$ in the actual medium and with the reflection response at the surface Wapenaar et al. (2014a). The focusing functions have a focal point in the subsurface at $\mathbf{x}_A$. This focal point serves as a virtual source for the Green's function. The $+$ and $-$ superscripts of the decomposed Green's function refer to the direction of propagation ($+$ for down and $-$ for up) from the virtual source at $\mathbf{x}_A$. The leftmost superscript indicates an up-$(-)$ or downward$(+)$ propagating field at the receiver locations. The relation between the two unknown focusing functions and the two unknown Green's functions is given by the following two equations (Wapenaar et al. (2021));

$$G^{-,+}(\mathbf{x}_R, \mathbf{x}_A, t) + f_1^-(\mathbf{x}_R, \mathbf{x}_A, t) = \int_{\mathbb{S}_0} \int_{t'=0}^{\infty} R(\mathbf{x}_R, \mathbf{x}_S, t') f_1^+(\mathbf{x}_S, \mathbf{x}_A, t - t') \mathrm{d}t' \mathrm{d}\mathbf{x}_S, \qquad (1)$$

$$G^{-,-}(\mathbf{x}_R, \mathbf{x}_A, t) + f_1^+(\mathbf{x}_R, \mathbf{x}_A, -t) = \int_{\mathbb{S}_0} \int_{t'=0}^{\infty} R(\mathbf{x}_R, \mathbf{x}_S, t') f_1^-(\mathbf{x}_S, \mathbf{x}_A, t' - t) \mathrm{d}t' \mathrm{d}\mathbf{x}_S. \qquad (2)$$

In the compact operator notation of van der Neut et al. (2015b) equations 1 and 2 are written as

$$G^{-,+} + f_1^- = Rf_1^+, \qquad (3)$$

$$G^{-,-} + f_1^{+\star} = Rf_1^{-\star}, \qquad (4)$$

where $\star$ denotes the time-reverse. These two equations contain four unknowns: two Green's functions and two focusing functions. The only known in these equations is the measured reflection response $R$. Wapenaar et al. (2013) use the reasoning that the Green's function and focusing function can, under certain circumstances, be separated in time. Therefore, a time window function (Wapenaar et al. (2014a)) $\Theta(t)$ is defined that passes the focusing function and removes the Green's function from the left-hand side of equations 3 and 4. For point-sources that radiate in all directions, one time window is sufficient for both the up- and downgoing traveling waves. Up- and downgoing plane-waves, on the other hand, propagate at opposite dipping angles; hence, two time windows are needed in the plane-wave algorithm. These time windows remove all events that arrive at later times than the direct wave traveling from the virtual source position $\mathbf{x}_A$ to the receiver position $\mathbf{x}_R$ at surface $\mathbb{S}_0$, including the direct wave itself. This results in the following two equations that only have two unknowns (assuming the direct arrival $f_{1,d}^+$ is known)

$$f_1^- = \Theta_b R f_1^+, \tag{5}$$

$$f_1^{+\star} - f_{1,d}^{+\star} = \Theta_a R f_1^{-\star}, \tag{6}$$

where $f_1^+ = f_{1,m}^+ + f_{1,d}^+$, with $f_{1,d}^+$ the direct arrival of $f_1^+$, and $f_{1,m}^+$ events that arrive before the direct arrival time $t_d$. The separation between $f_{1,d}^+$ and $f_{1,m}^+$ can be successfully applied when there are no overlapping reflection events with the direct response. In other cases, separation can still be applied, but requires additional steps to overcome interference (Zhang et al. (2019)). These time windows (Wapenaar et al. (2021)) are defined as

$$\Theta_b(t) = \theta(t_b - t), \tag{7}$$

$$\Theta_a(t) = \theta(t_a - t), \tag{8}$$

where $\theta(t)$ denotes a tapered Heaviside step function. Note that two time windows are defined: one at time $t_a$ and one at $t_b$. In the point-source algorithm, $t_b = t_d - \varepsilon = t_a$ which makes equation 8 equal to equation 7, with the window function $\theta(t_d - \varepsilon - t)$. The $\varepsilon$ takes into account the finite length of the band-limited wavelet and ensures that the direct wave is removed from the right-hand side of equation 6 (Broggini et al. (2014)). Epsilon is typically chosen as half the dominant wavelength.

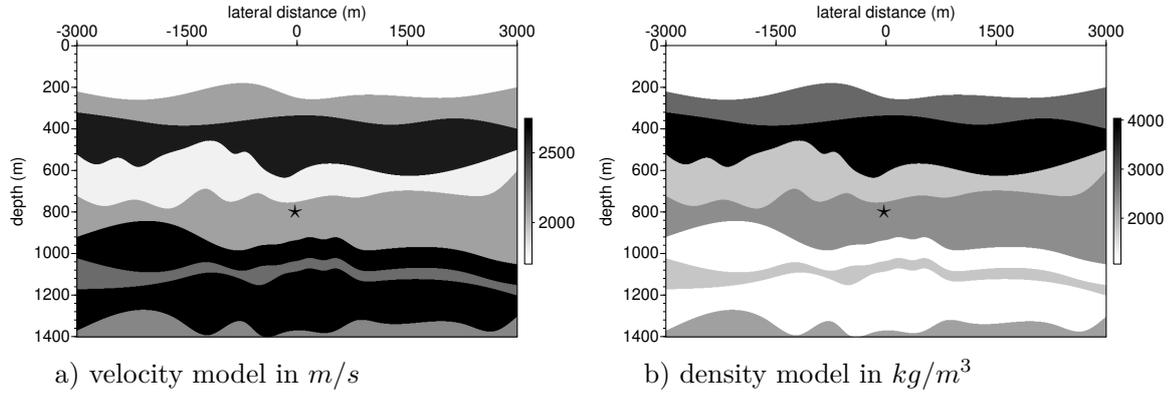

a) velocity model in $m/s$            b) density model in $kg/m^3$

Fig. 1: Multi layer model with velocity (a) and density (b) parameters. The location of the virtual point-source is marked with a $\star$.

To illustrate the application of the time windows, a virtual point-source is defined at a depth of 800 m in the laterally varying model of Figure 1 (after Meles et al. (2018)).

Figure 2 shows the focusing functions and Green's functions for the model of Figure 1, and the window functions (indicated with a dashed line) that separate these functions. Figure 2a represents the left-hand side of equation 3 and the time window that separates $f_1^-$ from $G^{-,+}$. In Figure 2b, the time window separates the time-reversal of $f_{1,m}^+$ from $G^{-,-}$ and represents the left-hand side of equation 4. The convolution/correlation in the right-hand sides of equations 3 and 4 is in practice carried out in the frequency domain, and a discrete Fourier transform in time is used to transform the sampled reflection data into the frequency domain. The discrete Fourier transform makes the fields periodic in time, with a periodicity equal to the number of time samples ($n_t$). Given this periodicity in time, reflection events occurring in time beyond $n_t$, end-up in negative times. The time windows defined in equations 7 and 8 pass all events earlier in time than $t = t_a$ and will also pass these time wrap-around events. To exclude these wrap-around events a time window is also implemented for negative times. In Figure 2a, we can see that the focusing functions also include events at negative times, and that these events are not present earlier than $-t_a = -t_d + \varepsilon$. Hence, the cutoff point of the time window at negative times is chosen at $-t_a$. The implemented time windows become

$$\Theta_b'(t) = \theta(t_b - t) - \theta(-t_b - t), \tag{9}$$

$$\Theta_a'(t) = \theta(t_a - t) - \theta(-t_a - t), \tag{10}$$

and the time windows at negative times, to suppress time wrap-around, are indicated with the dotted lines in Figure 2. There is no guarantee that this time window suppresses all wrap-around. If these windows are not sufficient to suppress the wrap-around, zeros can be padded to the reflection response.

To solve the unknown focusing functions in the coupled equations 5 and 6 different methods are developed. The iterative method described in Behura et al. (2014) and Wapenaar et al. (2014a) start with $f_{1,d}^+$ as the initial

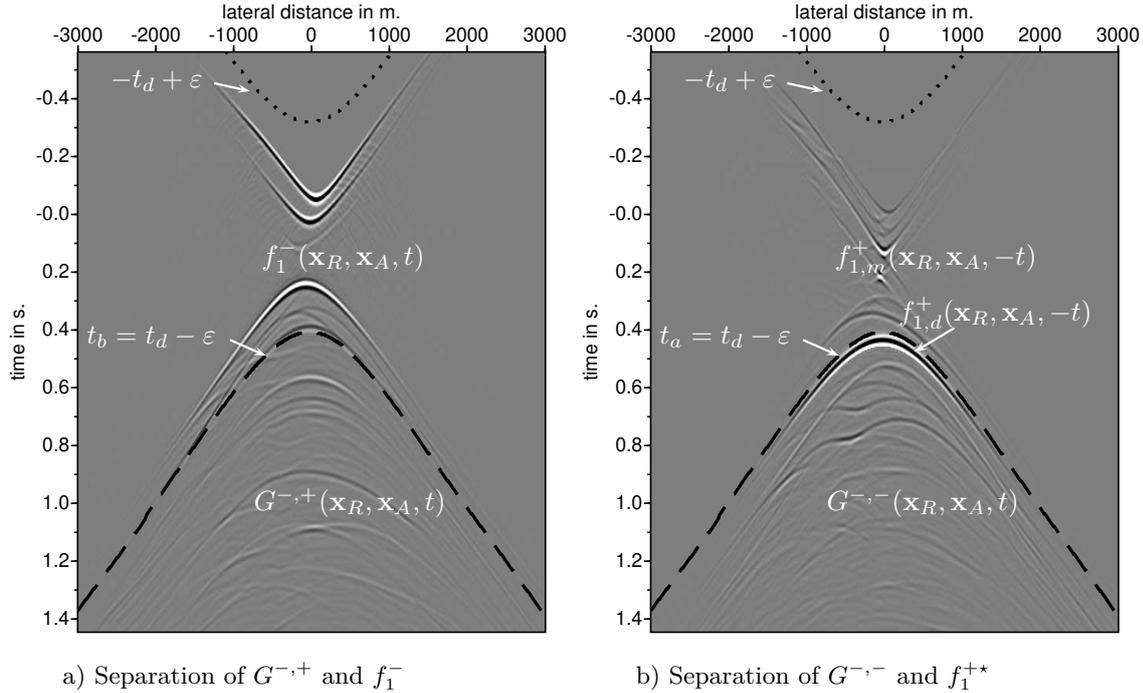

a) Separation of $G^{-,+}$ and $f_1^-$

b) Separation of $G^{-,-}$ and $f_1^{+\star}$

Fig. 2: Illustration of the time window function to separate the Green's function from the focusing function. The dashed black lines indicate the separation line of the time window and are indicated with white arrows. The dotted line indicates the time window that suppresses time wrap-around.

solution of $f_1^+$ and solve equation 5 for $f_1^-$ and substitute the solution in equation 6 to update $f_1^+$. This process is repeated until the updates to the focusing functions become very small. This iterative algorithm to solve the Marchenko equations is shown in Figure 3. The even iterations (starting the iteration count at 0 for the initial solution) in the scheme solve equation 5 and the odd iterations solve equation 6. Thorbecke et al. (2017) explain in more detail the implementation of this iterative algorithm.

Depending on the application it is not always needed to solve these equations until convergence. In Staring et al. (2018) the results of the first iteration are used to predict internal multiples and an adaptive subtraction method is used to suppress the predicted multiples. A direct least-squares inversion method to solve the coupled equations is discussed in van der Neut et al. (2015b) and Ravasi (2017).

Meles et al. (2018) show that plane-wave focusing functions $\tilde{f}_1^+$ and $\tilde{f}_1^-$ and associated plane-wave Green's functions $\tilde{G}^{-,+}$ and $\tilde{G}^{-,-}$ can be obtained by integrating an appropriate set of time-delayed focusing functions $f_1^+$ and $f_1^-$, each involving the solution of a Marchenko equation. The tildes represent plane-wave quantities for focusing functions and Green's functions. The plane-wave focusing functions (Wapenaar et al. (2021); Brackenhoff et al. (2022)) are defined by the following integration

$$\tilde{f}_1^\pm(\mathbf{x}, \mathbf{p}_A, t) = \int_{\mathbb{S}_A} f_1^\pm(\mathbf{x}, \mathbf{x}_A, t - \mathbf{p} \cdot \mathbf{x}_{H,A}) d\mathbf{x}_A,$$

(11)

with $\mathbf{p} = (p_1, p_2)$ and $p_1$ and $p_2$ horizontal ray parameters and $\mathbf{p}_A = (\mathbf{p}, x_{3,A})$ the ray parameter of the plane-wave at surface $\mathbb{S}_A$. The surface $\mathbb{S}_A$ is the depth level at which focusing takes place. The plane-wave Green's functions are defined by a similar integration. Here $\mathbf{x}_{H,A} = (x_{1,A} - x_{1,c}, x_{2,A} - x_{2,c})$, and $(x_{1,c}, x_{2,c})$ is the rotation point of the plane-wave. The rotation point is chosen in the center of the lateral extent of the plane-wave. This rotation point also defines the time origin $t = 0$ of the plane-wave. By making this choice for $t = 0$, the time-axis in the computed plane-wave Green and focusing functions is the same as in the point-source Marchenko solutions with a focal point at the rotation point of the plane-wave.

Note that the plane-wave integration in equation 11 for a time-reversed wave-field $P(\mathbf{x}, \mathbf{x}_A, -t)$ gives

$$\tilde{P}(\mathbf{x}, \mathbf{p}'_A, -t) = \int_{\mathbb{S}_A} P(\mathbf{x}, \mathbf{x}_A, -(t - \mathbf{p} \cdot \mathbf{x}_{H,A})) d\mathbf{x}_A,$$

(12)

with $\mathbf{p}'_A = (-\mathbf{p}, x_{3,A})$, a plane-wave dipping with the *opposite* angle as a plane-wave with $\mathbf{p}$. On a surface $\mathbb{S}_A$ with a homogeneous velocity along the surface, $\mathbf{p} \cdot \mathbf{x}_{H,A}$ are time shifts that are linearly proportional to the distance from the rotation point.

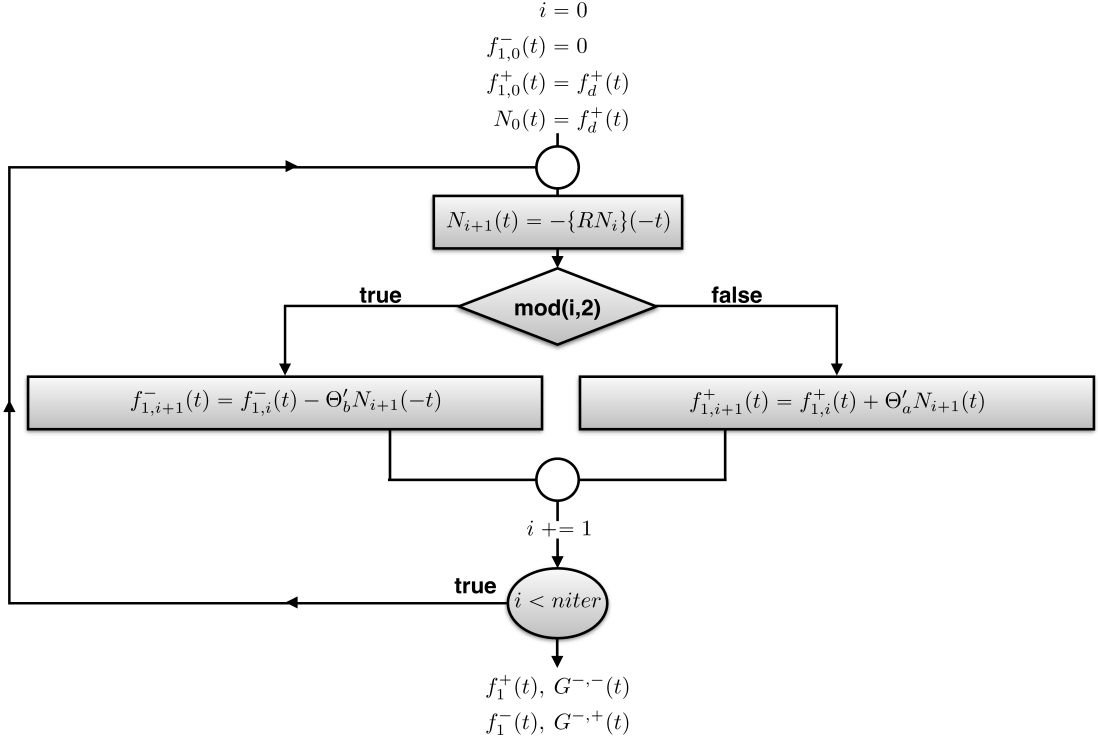

**Fig. 3:** Flow chart of the Marchenko algorithm. The down- ($f_1^+$) and up-going($f_1^-$) focusing functions are alternately updated. The scheme is finished after a pre-defined number of iterations and is usually chosen between 10-20 iterations.

Applying the same integration as in equation 11 over all fields results in the plane-wave representations of equations 3 and 4 (Meles et al. (2018))

$$\tilde{G}^{-,+}(\mathbf{x}_R, \mathbf{p}_A, t) + \tilde{f}_1^-(\mathbf{x}_R, \mathbf{p}_A, t) = \{R\tilde{f}_1^+\}(\mathbf{x}_R, \mathbf{p}_A, t), \tag{13}$$

$$\tilde{G}^{-,-}(\mathbf{x}_R, \mathbf{p}_A', t) + \tilde{f}_1^{+\star}(\mathbf{x}_R, \mathbf{p}_A, t) = \{R\tilde{f}_1^{-\star}\}(\mathbf{x}_R, \mathbf{p}_A, t). \tag{14}$$

Applying a time window that separates the Green function from the focusing function gives again two equations with two unknowns (assuming $\tilde{f}_{1,d}^{+\star}(\mathbf{x}_R, \mathbf{p}_A, t)$ is known);

$$\tilde{f}_1^-(\mathbf{x}_R, \mathbf{p}_A, t) = \tilde{\Theta}_b\{R\tilde{f}_1^+\}(\mathbf{x}_R, \mathbf{p}_A, t), \tag{15}$$

$$\tilde{f}_1^{+\star}(\mathbf{x}_R, \mathbf{p}_A, t) - \tilde{f}_{1,d}^{+\star}(\mathbf{x}_R, \mathbf{p}_A, t) = \tilde{\Theta}_a\{R\tilde{f}_1^{-\star}\}(\mathbf{x}_R, \mathbf{p}_A, t) \tag{16}$$

with $\tilde{f}_1^{+\star} = \tilde{f}_{1,m}^{+\star} + \tilde{f}_{1,d}^{+\star}$, where $\tilde{f}_{1,d}^{+\star}$ is the direct arrival of the plane-wave with a propagation angle defined by $(\mathbf{p}, x_{3,A})$, and $\tilde{f}_{1,m}^{+\star}$ contains the events that arrive before the direct arrival time $\tilde{t}_d$, where $\tilde{t}_d$ is the first arrival time of a plane-wave with a propagation angle defined by $(\mathbf{p}, x_{3,A})$. Note that the Green's functions in equations 13 and 14 contain plane-waves with opposite dipping angles defined by $\mathbf{p}_A$ and $\mathbf{p}_A'$. To separate $\tilde{G}^{-,+}(\mathbf{x}_R, \mathbf{p}_A, t)$ from $\tilde{f}_1^-(\mathbf{x}_R, \mathbf{p}_A, t)$ the direct arrival $\tilde{t}_d$, that is the first arrival time of $\tilde{G}^{-,+}(\mathbf{x}_R, \mathbf{p}_A, t)$, is required. The notation for $\tilde{t}_d$ or $\tilde{t}_d'$ is a choice, our choice is the same as made in Wapenaar et al. (2021). This choice uses the $'$ on $\tilde{t}_d$ the same as the $'$ on the propagation angle $\mathbf{p}_A$ in the upgoing Green's functions $\tilde{G}^{-,+}(\mathbf{x}_R, \mathbf{p}_A, t)$ and $\tilde{G}^{-,-}(\mathbf{x}_R, \mathbf{p}_A', t)$. Note that the defined arrival time $\tilde{t}_d'$ has the same arrival time of an upward propagating modeled plane-wave at $(x_{3,A})$ with propagation angle $\mathbf{p}_A'$.

The time window $\tilde{\Theta}_b(t)$ removes $\tilde{G}^{-,+}(\mathbf{x}_R, \mathbf{p}_A, t)$ from equation 13. The first non-zero contribution of $\tilde{G}^{-,+}(\mathbf{x}_R, \mathbf{p}_A, t)$ is at time $t_b = \tilde{t}_d - \varepsilon$. The time window $\tilde{\Theta}_a(t)$ removes $\tilde{G}^{-,-}(\mathbf{x}_R, \mathbf{p}_A', t)$ and $\tilde{f}_{1,d}^{+\star}(\mathbf{x}_R, \mathbf{p}_A, t)$ from equation 14, hence all events at times later than $t_a = \tilde{t}_d' - \varepsilon$ are set to zero. Similar to the point-source scheme these time windows are implemented with an additional window at negative times to suppress time wrap-around and are given by

$$\tilde{\Theta}_b'(t) = \theta(\tilde{t}_d - \varepsilon - t) - \theta(-\tilde{t}_d' + \varepsilon - t), \tag{17}$$

$$\tilde{\Theta}_a'(t) = \theta(\tilde{t}_d' - \varepsilon - t) - \theta(-\tilde{t}_d + \varepsilon - t). \tag{18}$$

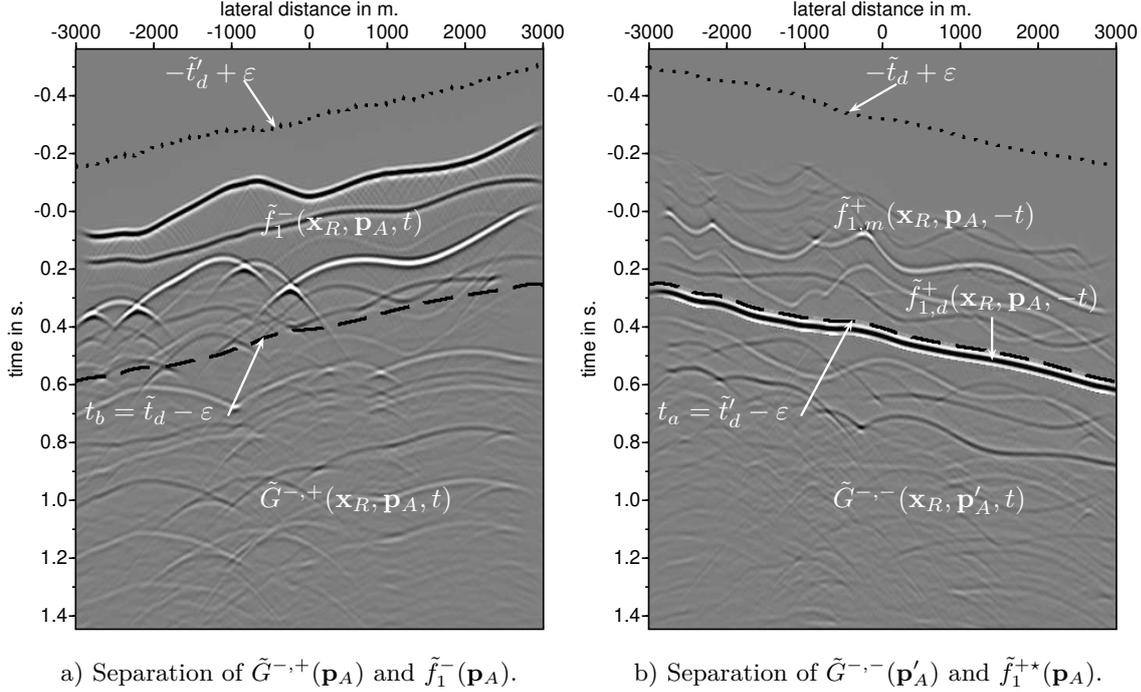

a) Separation of $\tilde{G}^{-,+}(\mathbf{p}_A)$ and $\tilde{f}_1^-(\mathbf{p}_A)$.

b) Separation of $\tilde{G}^{-,-}(\mathbf{p}'_A)$ and $\tilde{f}_1^{+\star}(\mathbf{p}_A)$.

**Fig. 4:** Illustration of the time window function to separate the plane-wave Green's function from the focusing function. The dashed black lines indicate the separation line of the time window and are indicated with white arrows. The dotted line indicates the time window that suppresses time wrap-around.

Similar to Figure 2, Figure 4 shows the plane-wave focusing functions and Green's functions, as in the left-hand side of equations 13 and 14, and the window functions separating them. In this example the depth of the plane-waves is chosen at 800 m in the model shown in Figure 1. In the following section these two time window functions are discussed in more detail and the differences with the point-source implementation of the Marchenko algorithm are explained.

### Basic algorithm

The plane-wave method is illustrated with two numerical examples, a laterally invariant and laterally variant medium. In two dimensions the downward propagating plane-wave focusing function $\tilde{f}_1^+(\mathbf{x}, \mathbf{p}_A, t)$ in equation 11, defines a plane-wave at the focal plane $\mathbb{S}_A$ with a dip angle $\alpha$ and $p_1 = \frac{\sin(\alpha)}{c}$. In the first numerical example we assume a medium with a laterally invariant velocity $c$ for each depth and shown in Figure 5a and b. Figure 6 shows $\tilde{f}_1^+(\mathbf{x}, \mathbf{p}_A, t)$, for a focal plane at a depth of 900 m at two different angles: Figure 6a with an angle of 0 degrees and Figure 6c with an angle of 3 degrees. In Figure 6b and 6d the focus function is shown for receivers at the focal level (900 m depth). Similar to the focal-point Marchenko method, the medium for the plane-wave focusing functions is chosen homogeneous below the focal level. The focus function $\tilde{f}_1^+(\mathbf{x}, \mathbf{p}_A, t)$ has a focus in time at $t = \mathbf{p} \cdot \mathbf{x}_{H,A}$. For a dipping plane-wave this time focus occurs at different positions in the model. Figure 6a and c shows the computed $\tilde{f}_1^+(\mathbf{x}, \mathbf{p}_A, t)$ at the surface (0 m depth) and includes an extra event around -0.15 seconds that compensates for the multiples generated between the first and second reflector at 400 and 700 meter respectively. The compensation of multiples for point-sources is explained in de detail in Zhang and Slob (2019). The focus functions at focal level (Figure 6b and d) show only one event, the downgoing event present at the surface is compensated by the reflected event from the reflector at 700 meter depth.

To illustrate this compensation effect, snapshots of the focusing function $\tilde{f}_1^+(\mathbf{x}, \mathbf{p}_A, t)$ (Figure 6a and 6c) propagating into the truncated medium (that is homogeneous below the last reflector at 700 m depth) are shown in Figure 7 for the same angles of 0 and 3 degrees. Figure 7a to 7d shows four different snapshots of the superposition of the down-going horizontal plane-wave $\tilde{f}_1^+(\mathbf{x}, \mathbf{p}_A, t)$ and upgoing plane-wave $\tilde{f}_1^-(\mathbf{x}, \mathbf{p}_A, t)$. The snapshots of a plane-wave with an angle of 3 degrees are shown in Figure 7e to 7h. At 0.05 seconds before t=0 (Figure 7a and 7e), there are two upward traveling reflected waves from the interfaces at 400 and 700 meter depth and two downgoing events from $\tilde{f}_1^+(\mathbf{x}_R, \mathbf{p}_A, t)$. The snapshots in Figure 7f (at $t = 0.00$ s), 7c and g (at $t = +0.05$ s) show that the second downward traveling event coincides at the first interface (at 400 m depth) with the upgoing reflection of the second interface (at 700 m depth) and these events compensate each other.

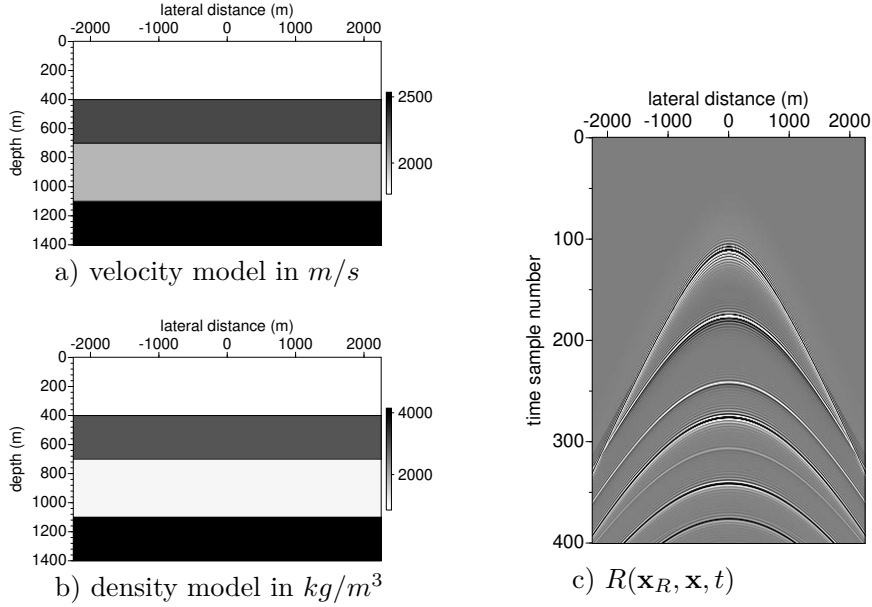

a) velocity model in $m/s$

b) density model in $kg/m^3$

c) $R(\mathbf{x}_R, \mathbf{x}, t)$

Fig. 5: Two dimensional four layer model with velocity (a) and density (b) parameters. A common-source record, with source position $\mathbf{x} = (x_1 = 0, x_3 = 0)$ and receivers at $\mathbf{x}_R = (x_1 = x_R, x_3 = 0)$ (c). The source wavelet in $R$ has a flat frequency spectrum from 5 to 90 Hz.

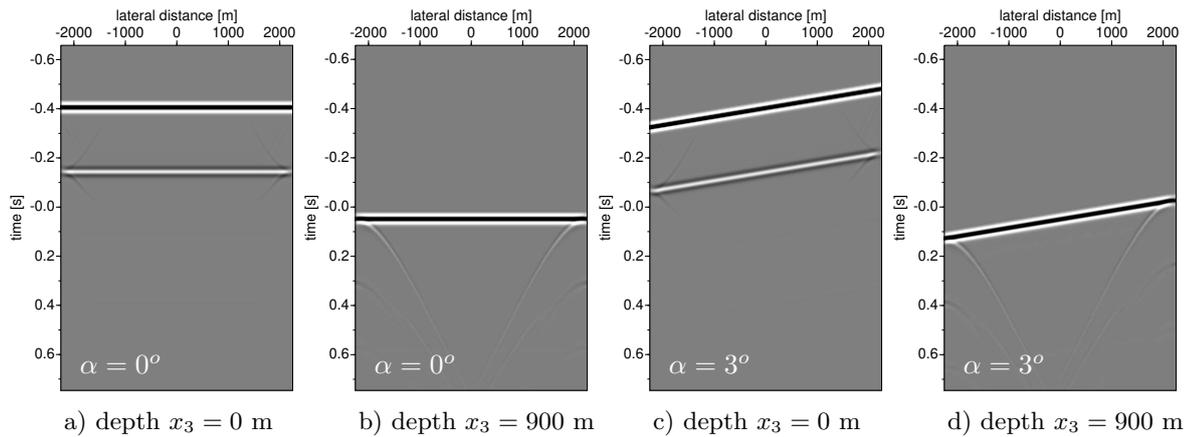

a) depth $x_3 = 0$ m     b) depth $x_3 = 900$ m     c) depth $x_3 = 0$ m     d) depth $x_3 = 900$ m

Fig. 6: Time recordings of the plane-wave focusing function $\tilde{f}_1^+(\mathbf{x}_R, \mathbf{p}_A, t)$ with a focal depth of 900 meter measured with receivers at $x_3 = 0$ and $x_3 = 900$ meter in the truncated medium for two different plane-wave propagation angles (0 and 3 degrees). At the end-points of the plane-wave, diffraction curves are present due to the limited lateral extent of the constructed plane-wave.

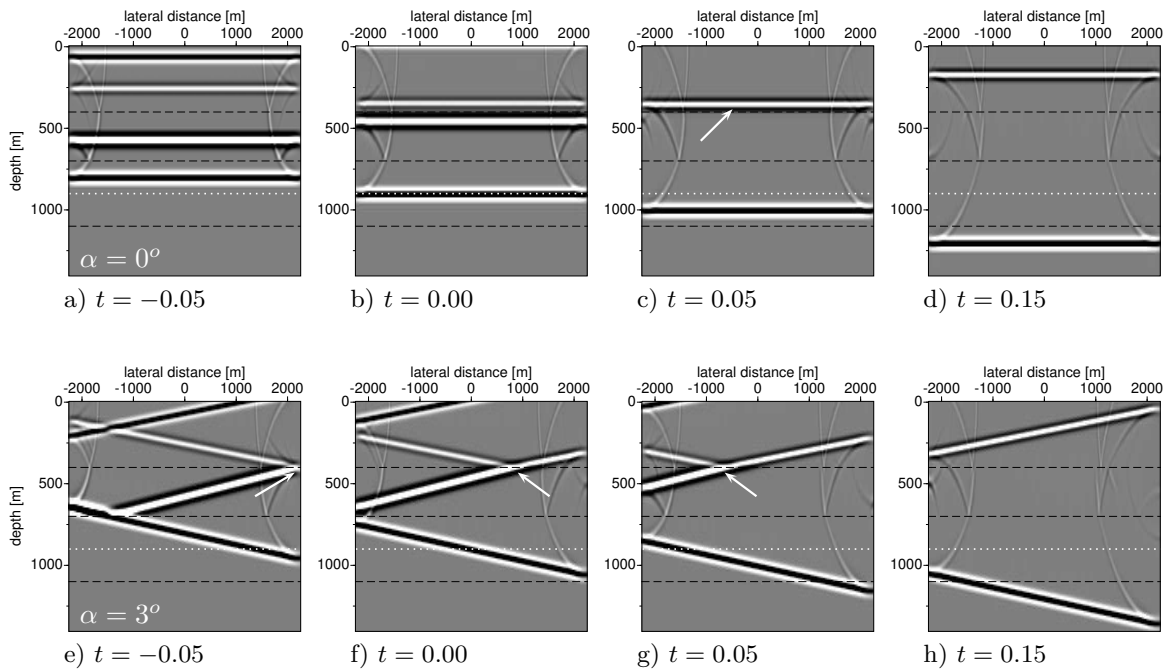

**Fig. 7:** Time snapshots for a propagating $\tilde{f}_1^+(\mathbf{x}_R, \mathbf{p}_A, t) + \tilde{f}_1^-(\mathbf{x}_R, \mathbf{p}_A, t)$ at two different plane-wave propagation angles. Note the diffraction effects at the edges of the plane-wave. The white dotted-line indicates the focal depth of the plane-wave.

This is indicated with an arrow in the pictures. The fourth snapshot show that after this compensation all the internal multiples between the reflectors at 400 and 700 m depth have vanished and only one downgoing direct wave and an upgoing reflected wavefield (from the reflector at 700 m depth) are remaining. The compensation of the first upward traveling multiple indicates that $\tilde{f}_1^+(\mathbf{x}_R, \mathbf{p}_A, t)$ is a solution of the Marchenko equations. The illustration in Figure 7 demonstrates that the internal multiple compensation principle also holds for the plane-wave Marchenko method.

In Figure 8 the experiment is repeated in the laterally variant medium of Figure 1. The focal-plane is chosen at 800 m depth, just below the fourth reflector. The same observation is made; the downgoing events in $\tilde{f}_1^+(\mathbf{x}_R, \mathbf{p}_A, t)$ compensate the upgoing events at interfaces and suppress the generation of internal multiples.

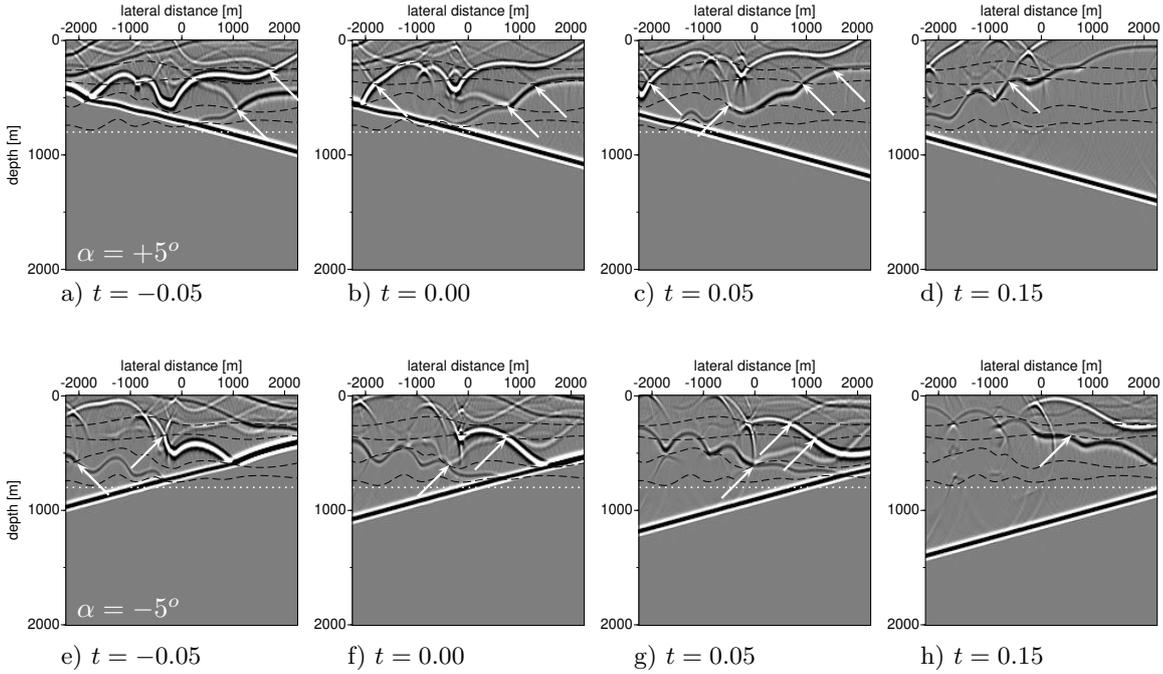

**Fig. 8:** Time snapshots for a propagating $\tilde{f}_1^+(\mathbf{x}_R, \mathbf{p}_A, t) + \tilde{f}_1^-(\mathbf{x}_R, \mathbf{p}_A, t)$ in model of Figure 1 for two different plane-wave propagation angles. Note that there are remaining diffraction effects originating from edges on the interfaces. The white dotted-line indicates the focal depth of the plane-wave. The arrows indicate positions at a reflector where an up-going reflected field, that generates internal multiples, is compensated by a down-going event from the focusing function.

### Horizontal plane-waves

To start the iterative Marchenko algorithm for plane-waves $\tilde{f}_1^+(\mathbf{x}_R, \mathbf{p}_A, t)$ is required; the first arrival of a plane-wave response from a focal-plane in the subsurface. This forward modeling step can be computed in a macro model estimated from the reflection data. The computed initial response is then muted below the first arrival times to get the time-reversal of the initial focusing field $\tilde{f}_{1,d}^+(\mathbf{x}_R, \mathbf{p}_A, t)$. A first example is made for a horizontal (zero-degree) plane-wave defined at 800 meter depth in the laterally varying model of Figure 1 (same model as in Meles et al. (2018)).

Figure 9a shows the forward modeled plane-wave response of a horizontal plane-wave at 800 m depth and receivers at the surface. The first arrivals of that plane-wave, shown in Figure 9b, is the time-reversal of the input field $\tilde{f}_{1,d}^+(\mathbf{x}_R, \mathbf{p}_A, t)$ of the Marchenko scheme. The computed down- and up-going Green's functions, after 16 iterations of equations 15 and 16 and substituting the results in equations 13 and 14, are shown in 9c and 9d respectively and give the expected response. The horizontal plane-wave Marchenko mute-line (as defined by $\tilde{\Theta}'_{a,b}$ in equations 17 and 18), to separate the Green's function from the focusing functions, is symmetric around $t = 0$ because $\tilde{t}_d = \tilde{t}'_d$. This is the same time symmetry as in use for the Marchenko point-source algorithm.

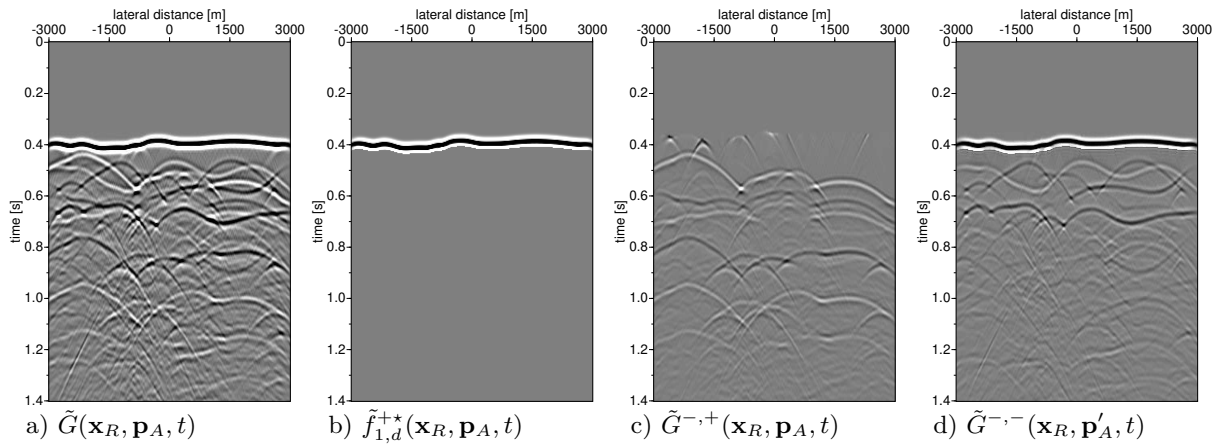

**Fig. 9:** Results of the plane-wave Marchenko scheme for a horizontal plane-wave $\mathbf{p}_A = (\mathbf{0}, x_{3,A})$. Adding the up- and downgoing Green's functions of c and d, that are computed with the Marchenko algorithm, gives the same wavefield as the directly forward modeled result in a. All figures are plotted with the same clipping factor.

## Dipping plane-waves

The Marchenko algorithm for dipping plane-waves follows the same procedure as for horizontal plane-waves. As indicated by equations 17 and 18 the Marchenko time windows have to be designed differently for dipping plane-waves. For horizontal plane-waves the implemented time window is symmetric around $t = 0$ and $\tilde{\Theta}'_a = \tilde{\Theta}'_b$. This does not hold anymore for dipping plane-waves. Figure 10 is an example of time windows that are designed for a dipping plane-wave of $+5$ degrees. Two time windows are designed: one for the even iterations (Figure 10a), in use by equation 15, and one for the odd iterations, that has a reverse angle (Figure 10b) and in use by equation 16. To design these windows for a positive angle, the first arrival time for a plane-wave at the same depth level with a reverse angle is needed as well.

In the following we explain in more detail what is expressed in the plane-wave Marchenko equations 15 and 16. The plane-wave Marchenko scheme starts with forward modeling the response to a (dipping) plane-wave at depth. This modeled wave-field response is given in Figure 12a, where the depth of the plane-wave is chosen at 800 m. The direct arrival is selected from this modeled field and shown in Figure 12b. The time reverse of this field will be $\tilde{f}^+_{1,d}(\mathbf{x}_R, \mathbf{p}_A, t)$. This field $\tilde{f}^+_{1,d}(\mathbf{x}_R, \mathbf{p}_A, t)$ is, together with the reflection data $R$, input of the plane-wave Marchenko scheme.

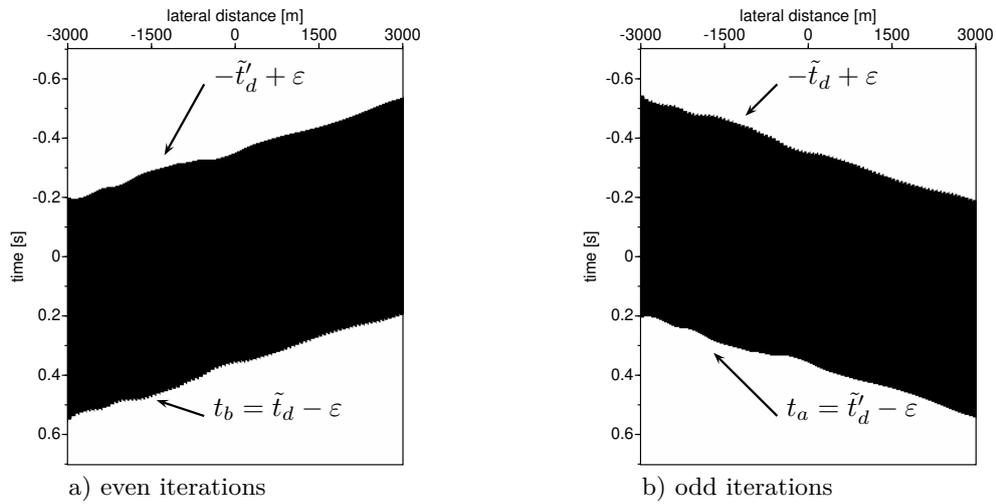

**Fig. 10:** The time windows for dipping plane-waves for even (a) and odd (b) iterations for an angle of +5 degrees. The wavefields in the black area of the windows pass, the fields in the white area are set zo zero.

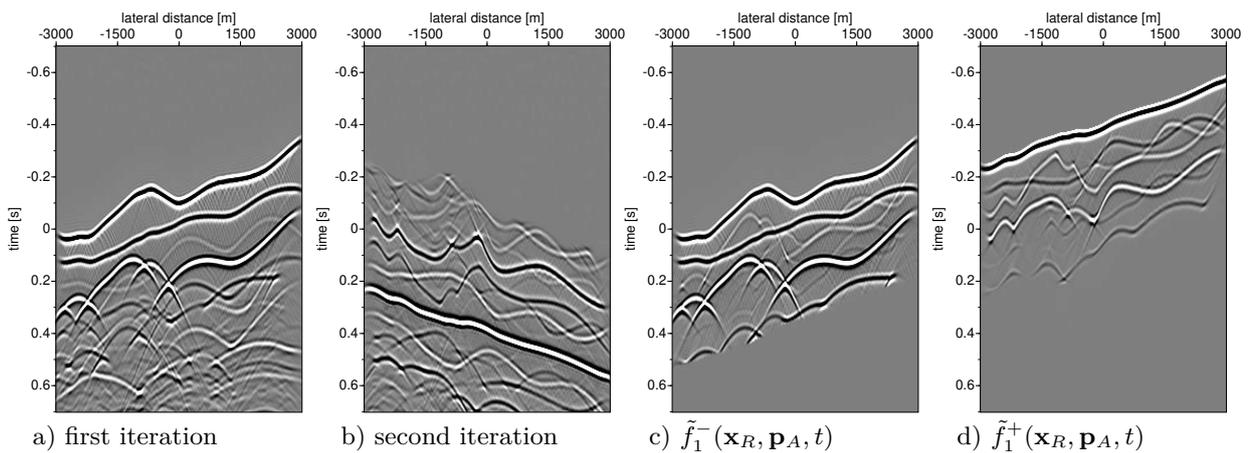

**Fig. 11:** Basic plane-wave Marchenko results for a plane-wave with an angle of incidence of +5 degrees. Note, that the results of the first iteration (a) is dipping in the opposite direction as the second iteration (b) and the algorithm uses the time windows, designed for dipping plane-waves as given in equations 17 and 18, to take this into account.

In the Marchenko algorithm the iterations are alternating between a convolution of the focusing functions with $R$, or a correlation with $R$ (Thorbecke et al. (2017)). In the first step, correlation by equation 16, the wavefield is shifted backward in time related to the times of $\tilde{t}_d$, and in the second step, convolution by equation 15, the wavefield is shifted forward in time related to the times of $\tilde{t}'_d$. In Figure 11a the result of the first iteration (correlation) is shown and in Figure 11b the result of the second iteration (convolution) is shown. In Figure 11a we can see that the first event, that starts at negative time, shows an imprint of the undulation of the first reflector and has an opposite dip compared to the plane-wave in Figure 12a.

The result of the first iteration in Figure 11a is windowed in time (with the window in Figure 10a) to mute $\tilde{G}^{-,+}(\mathbf{x}_R, \mathbf{p}_A, t)$, followed by time-reversal and convolution with $R$ (equation 16). The result is shown in Figure 11b. In this second iteration the convolution step brought the data back to the arrival times corresponding to reflection times in the forward modeled plane-wave response of Figure 12a. Note that the arrival time in Figure 11b, starting from the left at time 0.2 s dipping to the right to 0.5 s, is the same as the first arrival time $\tilde{t}_d$ in Figure 12a and 12b. The result of this second iteration is muted in time (with the window in Figure 10b) to remove $\tilde{G}^{-,-}(\mathbf{x}_R, \mathbf{p}'_A, t)$ and the first arrival event at $\tilde{f}^+_{1,d}(\mathbf{x}_R, \mathbf{p}_A, t)$, as indicated in Figure 4b. The odd iteration(s) are building up $\tilde{f}^-_1(\mathbf{x}_R, \mathbf{p}_A, t)$ and $\tilde{G}^{-,+}(\mathbf{x}_R, \mathbf{p}_A, t)$ and the even iteration(s) are building up $\tilde{f}^+_1(\mathbf{x}_R, \mathbf{p}_A, t)$ and $\tilde{G}^{-,-}(\mathbf{x}_R, \mathbf{p}'_A, t)$. Figure 11c and 11d show $\tilde{f}^+_1(\mathbf{x}_R, \mathbf{p}_A, t)$ and $\tilde{f}^-_1(\mathbf{x}_R, \mathbf{p}_A, t)$ respectively after 16 iterations.

In Figure 12 three plane-wave responses are shown with angles of -5, 0 and 5 degrees after 16 iterations. Comparing these three plane-wave responses for $\tilde{G}^{-,-}$ and $\tilde{G}^{-,+}$ shows that each angle illuminates different parts of the medium. This is clearly seen in the events that arrive later than 1.2 seconds. By combining different plane-wave responses into one image a fully illuminated image can be constructed by using only a few migrations (Meles et al. (2018)). Plane-wave imaging, that suppress internal multiples, can use the same strategies as point-source Marchenko, for example double focusing as described in van der Neut et al. (2017); Staring et al. (2018), or Multi Dimensional Deconvolution as discussed in Ravasi et al. (2016). Almobarak (2021) discusses different plane-wave imaging methods and shows that the Marchenko Green's function plane-wave response is a computational efficient imaging method.

Figure 13 shows horizontal plane-wave images from the Troll field data-set located west of Norway, that was kindly provided by Equinor. This data set is part of a time-lapse monitoring set. We have selected a small part of this data-set, with source and receiver positions on the same locations. The receiver and source spacing is 12.5 meter and traces are recorded with a time sampling of 4 ms. The data-set has been pre-processed to remove free-surface multiples and deconvolve the wavelet (Qu and Verschuur (2020)).

The imaging is carried out according to the imaging method described in Meles et al. (2018). In the basic imaging method, a forward modeled plane-wave response is computed at each depth level in an estimated smooth macro model of the data. This plane-wave depth response is correlated with the point-source responses of the recorded data and integrated over the receiver coordinate for each point-source response. This creates the plane-wave depth response of the data (Rietveld et al. (1992)). This plane-wave response of the data is correlated with the same modeled plane-wave response at each depth level and the imaging condition $t = 0$ is used to construct the image for all depth levels. Combining these depth levels gives the left side picture in Figure 13 labeled 'standard'.

To compute the Marchenko based image, the first arrivals of the forward modeled plane-wave response at each depth level is input to the plane-wave Marchenko algorithm to create $\tilde{G}^{-,+}$. The computed $\tilde{G}^{-,+}$ is, similar to the standard imaging method, correlated with the forward modeled plane-wave response for each depth level, and the imaging condition at $t = 0$ constructs the image for all depth levels. For a horizontal plane-wave ($\mathbf{p}_A = (\mathbf{p} = \mathbf{0}, x_{3,A})$) at one depth level $x_{3,A}$ this imaging condition is represented by

$$\tilde{I}(\mathbf{x}_R, \mathbf{p}_A) = \int_t \tilde{f}^+_{1,d}(\mathbf{x}_R, \mathbf{p}_A, t)\tilde{G}^{-,+}(\mathbf{x}_R, \mathbf{p}_A, t)dt. \qquad (19)$$

The advantage of using $\tilde{G}^{-,+}$ is that this field does not contain downgoing internal multiples from the layers above the focal/imaging depth. The Marchenko method has separated the internal multiples in up- and downgoing parts in the Greens functions by applying the computed focal functions to the reflection data. Alternative strategies to compute an image without internal multiples can be based on MME. Thorbecke et al. (2021) illustrate the working of the internal multiple elimination of MME directly on reflection data in Figure 9 and equations (12)-(20). Staring et al. (2018) illustrate source-receiver Marchenko redatuming and imaging on field data using an adaptive double-focusing method. Figure 4 of that paper shows internal multiples that are first predicted and then subtracted from the data.

The middle picture of Figure 13 shows the Marchenko-created image. The difference between the standard image and the Marchenko-based image is shown in Figure 13c. From this difference plot it is observed that the Marchenko method predicts and removes internal multiples. In this data-set the effect of the multiple removal on the image is small. van IJsseldijk et al. (2023) show that Marchenko multiple removal on this data set improves the confidence of the effects of small time-lapse changes.

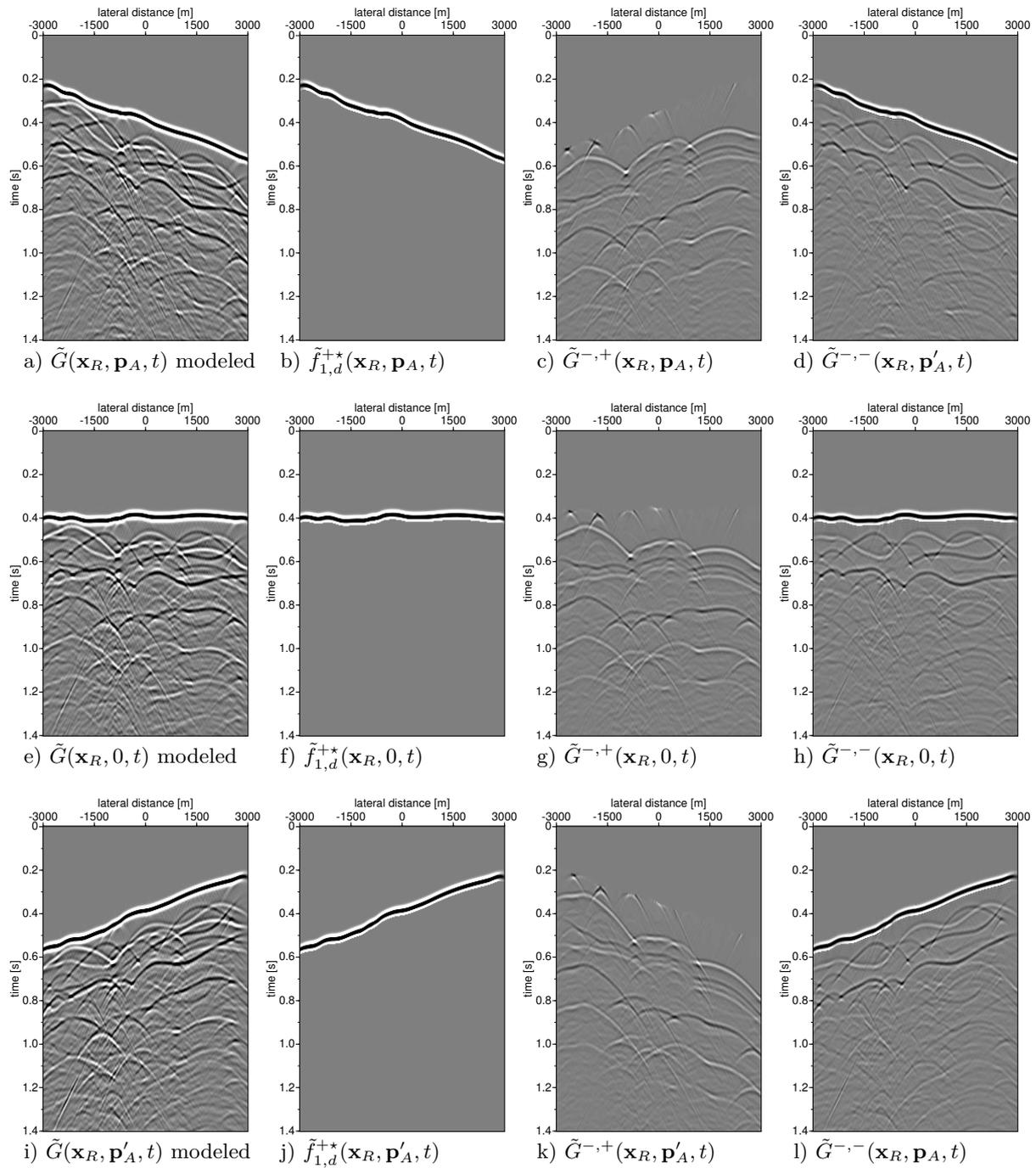

Fig. 12: Marchenko computed plane-wave responses for angles at 5 (a-d), 0 (e-h), and -5 (i-l) degrees. Note the difference in illumination in the decomposed Green's function for different dipping angles of the plane-wave. Analog to Figure 9 the addition of the Marchenko computed up- (d) and downgoing (k) Green's function gives the forward modeled response in a.

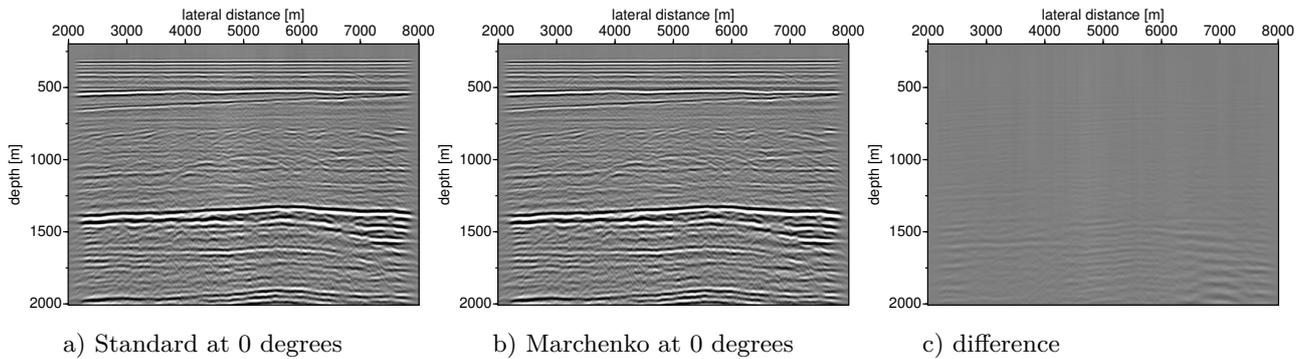

a) Standard at 0 degrees     b) Marchenko at 0 degrees     c) difference

**Fig. 13:** Plane-wave images of the Troll field data-set for a horizontal plane-wave using standard imaging (left) and Marchenko based imaging (middle). All images are displayed with the same clipping factor.

## Conclusions

The plane-wave Marchenko method is a straightforward extension of the point-source Marchenko method. A counter-intuitive aspect of the plane-wave method is that the retrieved up- and down-going Marchenko Green's functions have opposite dipping angles. This is taken into account in the time windows that separate the Green's function from the focusing function. In case one would like to get the total Green's function for a specific dip angle, one would have to run the whole procedure twice, for opposite dip angles. In this paper, the use of these time windows is illustrated with numerical and field data examples. The plane-wave Marchenko method can give a computational advantage over the point-source method. Specially for imaging applications with 3-dimensional data sets that have moderate lateral changes, in that case only a few plane-wave migrations are needed to compute a well illuminated image.

## Acknowledgment

The authors thank Equinor (formerly Statoil A.S.) for providing the marine dataset of the Troll Field. This research was funded by the European Research Council (ERC) under the European Union's Horizon 2020 research and innovation program (grant agreement no. 742703).

## Code availability

- Name of the code/library: OpenSource code for Finite Difference, Marchenko algorithms and processing utilities

- Contact: j.w.thorbecke@tudelft.nl (+31622211997)

- Hardware requirements: tested on x86_64 and aarch64 processors

- Program language: C and Fortran

- Software required: C compiler, Fortran compiler, GNU Make, tested only on Linux/Unix OS. The display and generation of the figures is done with Seismic Unix and is available at https://github.com/JohnWStockwellJr/SeisUnix.git.

- Program size: 147 MB

  The source codes are available for downloading at the link: https://gitlab.com/geophysicsdelft/OpenSource.git The scripts to reproduce the results in this manuscript can be found in `.../OpenSource/marchenko/demo/planewave`. The README in that directory explains all the steps to reproduce the results in the manuscript. For the reproduction of the measured data example please contact us directly, we will ask the owner of the data if we can share the data.

## References

Almobarak, M., 2021. Plane-wave Marchenko imaging method: applications. Master's thesis. Delft University of Technology.

Behura, J., Wapenaar, K., Snieder, R., 2014. Autofocus imaging: Image reconstruction based on inverse scattering theory. Geophysics 79, A19 – A26.

Brackenhoff, J., Thorbecke, J., Meles, G.A., Koehne, V., Barrera, D.F., Wapenaar, K., 2022. 3d marchenko applications: implementation and examples. Geophysical Prospecting 70, 35 – 36.

Brackenhoff, J., Thorbecke, J., Wapenaar, K., 2019. Virtual sources and receivers in the real earth: Considerations for practical applications. Journal of Geophysical Research: Solid Earth 124, 802 – 821.

Broggini, F., Wapenaar, K., van der Neut, J., Snieder, R.K., 2014. Data-driven green's function retrieval and application to imaging with multidimensional deconvolution. Journal of Geophysical Research: Solid Earth 119, 425 – 441.

van IJsseldijk, J., Brackenhoff, J., Thorbecke, J., Wapenaar, K., 2023. Time-lapse applications of the marchenko method on the troll field. Geophysical Prospecting under review, –.

van IJsseldijk, J., van der Neut, J., Thorbecke, J., Wapenaar, K., 2022. Extracting small time-lapse traveltime changes in a reservoir using primaries and internal multiples after marchenko-based target zone isolation. Geophysics 88, R135 – R143.

Meles, G.A., Wapenaar, K., Thorbecke, J., 2018. Virtual plane-wave imaging via marchenko redatuming. Geophysical Journal International 214, 508 – 519.

Meles, G.A., Zhang, L., Thorbecke, J., Wapenaar, K., Slob, E.C., 2020. Data-driven retrieval of primary plane-wave responses. Geophysical Prospecting 68, 1834 – 1846.

van der Neut, J., Johnson, J.L., van Wijk, K., Singh, S., Slob, E.C., Wapenaar, K., 2017. A marchenko equation for acoustic inverse source problems. The Journal of the Acoustical Society of America 141, 4332 – 4346.

van der Neut, J., Thorbecke, J., Wapenaar, K., Slob, E., 2015a. Inversion of the multidimensional marchenko equation, in: 77th Annual International Meeting, Extended Abstracts, European Association of Geoscientists and Engineers. pp. We–N106–04.

van der Neut, J., Vasconcelos, I., Wapenaar, K., 2015b. On green's function retrieval by iterative substitution of the coupled marchenko equations. Geophysical Journal International 203, 792 – 813.

Qu, S., Verschuur, E., 2020. Simultaneous joint migration inversion for high-resolution imaging/inversion of time-lapse seismic datasets. Geophysical Prospecting 68, 1167 – 1188.

Ravasi, M., 2017. Rayleigh-marchenko redatuming for target-oriented, true-amplitude imaging. Geophysics 82, S439 – S452.

Ravasi, M., Vasconcelos, I., Kritski, A., Curtis, A., da Costa Filho, C., Meles, G.A., 2016. Target-oriented marchenko imaging of a north sea field. Geophysical Journal International 205, 99 – 104.

Rietveld, W., Berkhout, A., Wapenaar, C., 1992. Optimum seismic illumination of hydrocarbon reservoirs. Geophysics 57, 1334 – 1345.

Slob, E.C., Wapenaar, K., Broggini, F., Snieder, R.K., 2014. Seismic reflector imaging using internal multiples with marchenko-type equations. Geophysics 79, S63 – S76.

Staring, M., Pereira, R., Douma, H., van der Neut, J., Wapenaar, K., 2018. Source-receiver marchenko redatuming on field data using an adaptive double-focusing method. Geophysics 83, S579 – S590.

Stockwell, J.W., Cohen, J.K., 2016. Cwp/su: Seismic unix package. URL: `https://github.com/JohnWStockwellJr/SeisUnix.git`.

Thorbecke, J., Brackenhoff, J., 2023. Opensource code for finite difference, marchenko algorithms and processing utilities. URL: `https://gitlab.com/geophysicsdelft/OpenSource.git`.

Thorbecke, J., Slob, E., Brackenhoff, J., van der Neut, J., Wapenaar, K., 2017. Implementation of the marchenko method. Geophysics 82, WB29 – WB45.

Thorbecke, J., Zhang, L., Wapenaar, K., Slob, E., 2021. Implementation of the marchenko multiple elimination algorithm. Geophysics 86, F9 – F23.

Wapenaar, K., Brackenhoff, J., Dukalski, M., Meles, G.A., Slob, E., Staring, M., Thorbecke, J., van der Neut, J., Zhang, L., Urruticoechea, C.R., 2021. Marchenko redatuming, imaging and multiple elimination, and their mutual relations. Geophysics 86, WC117 – WC140.

Wapenaar, K., Broggini, F., Slob, E.C., Snieder, R.K., 2013. Three-dimensional single-sided marchenko inverse scattering, data-driven focusing, green's function retrieval, and their mutual relations. Physical review letters 110, 084301.

Wapenaar, K., Broggini, F., Snieder, R.K., 2012. Creating a virtual source inside a medium from reflection data: heuristic derivation and stationary-phase analysis. Geophysical Journal International 190, 1020 – 1024.

Wapenaar, K., Thorbecke, J., van der Neut, J., Broggini, F., Slob, E., Snieder, R., 2014a. Marchenko imaging. Geophysics 79, WA39 – WA57.

Wapenaar, K., Thorbecke, J., van der Neut, J., Broggini, F., Slob, E.C., Snieder, R.K., 2014b. Green's function retrieval from reflection data, in absence of a receiver at the virtual source position. The Journal of the Acoustical Society of America 135, 2847 – 2861.

Zhang, L., Slob, E., 2020. A fast algorithm for multiple elimination and transmission compensation in primary reflections. Geophysical Journal International 221, 371 – 377.

Zhang, L., Slob, E.C., 2019. Free-surface and internal multiple elimination in one step without adaptive subtraction. Geophysics 84, A7 – A11.

Zhang, L., Thorbecke, J., Wapenaar, K., Slob, E., 2019. Transmission compensated primary reflection retrieval in the data domain and consequences for imaging. Geophysics 84, Q21 – Q36.